\documentclass[Afour,sageh,times]{sagej}

\usepackage{moreverb}
\usepackage{xurl}
\usepackage{pbox}

\usepackage{graphicx}
\usepackage{color}
\usepackage{multirow}
\usepackage{tikz}
\usetikzlibrary{arrows.meta,bending,matrix}
\usepackage{subcaption}
\usepackage{algorithm}
\usepackage[noend]{algpseudocode}
\algnewcommand\And{\textbf{ and }}
\algnewcommand\Not{\textbf{not }}
\usepackage{nameref}
\usepackage{flushend}
\usepackage{wrapfig}
\usepackage{balance}
\MakeRobust{\Call}

\usepackage[colorlinks,bookmarksopen,bookmarksnumbered,citecolor=red,urlcolor=red]{hyperref}
\newcommand\BibTeX{{\rmfamily B\kern-.05em \textsc{i\kern-.025em b}\kern-.08em
T\kern-.1667em\lower.7ex\hbox{E}\kern-.125emX}}

\def\volumeyear{2022}
\newcommand{\ts}{\textsuperscript}

\newcommand{\review}[1]{\textcolor{black}{#1}}

\begin{document}

\runninghead{S. Iserte \textsc{et al.}}

\title{A Study on the Performance of Distributed Training of Data-driven CFD Simulations}

\ifdefined\ANON
\author{Authors omitted for anonymized reviews}
\else
\author{Sergio Iserte\affilnum{1,2}, Alejandro González-Barberá\affilnum{1}, Paloma Barreda\affilnum{1}, and Krzysztof Rojek\affilnum{3}}
\affiliation{\affilnum{1}Universitat Jaume I, Spain\\
\affilnum{2}Barcelona Supercomputing Center, Spain\\
\affilnum{3}Czestochowa University of Technology, Poland}
\corrauth{S. Iserte, Universitat Jaume I, 12071 Castelló de la Plana, Spain.}
\email{siserte@uji.es}
\fi

\begin{abstract}
Data-driven methods for computer simulations are blooming in many scientific areas.
The traditional approach to simulating physical behaviors relies on solving partial differential equations (PDE). 
Since calculating these iterative equations is highly both computationally demanding and time-consuming, data-driven methods leverage artificial intelligence (AI) techniques to alleviate that workload.
Data-driven methods have to be trained in advance to provide their subsequent fast predictions, however, the cost of the training stage is non-negligible.

This paper presents a predictive model for inferencing future states of a specific fluid simulation that serves as a use case for evaluating different training alternatives.
Particularly, this study compares the performance of only CPU, multi-GPU, and distributed approaches for training a time series forecasting deep learning (DL) model.

With some slight code adaptations, results show and compare, in different implementations, the benefits of distributed GPU-enabled training for predicting high-accuracy states in a fraction of the time needed by the computational fluid dynamics (CFD) solver.
\end{abstract}


\maketitle

\section{Introduction}

Artificial intelligence (AI) methods have become pervasive in recent years due to numerous algorithmic advances and the accessibility of computational power~\cite{Dongarra,Domain}. 
In computational fluid dynamics (CFD), AI has been used to assist, accelerate, enhance, or even replace existing physical solvers based on partial differential equations (PDE)~\cite{Brunton2020}.
Especially, for the last few years, deep learning (DL) methods have been widely used in the field of CFD.

In this work, we propose a domain-specific AI model that can be easily integrated with CFD simulations.
The model training is thoroughly evaluated with distributed techniques on CPU and GPU devices.

The study presented throughout this paper is based on a homogenization tank. Homogenization tanks are essential in various industrial fields, especially those involving production processes where large volumes of liquids or aqueous mixtures are handled.

This type of tank is usually located in the stages prior to industrial mixing or deposition/fermentation treatments of different types of substances. Through this, as most industries have continuous operation processes working 24 hours a day, as well as, long waiting times between each phase, these facilities allow for more detailed control over the input flow into the following stage of the process, maintaining a regular input volume to ensure that the subsequent process has the necessary hydraulic retention time to carry out its function. 

\begin{figure}[tbp]
    \centering
    \includegraphics[trim={0 0 0 0}, clip, width=\linewidth]{"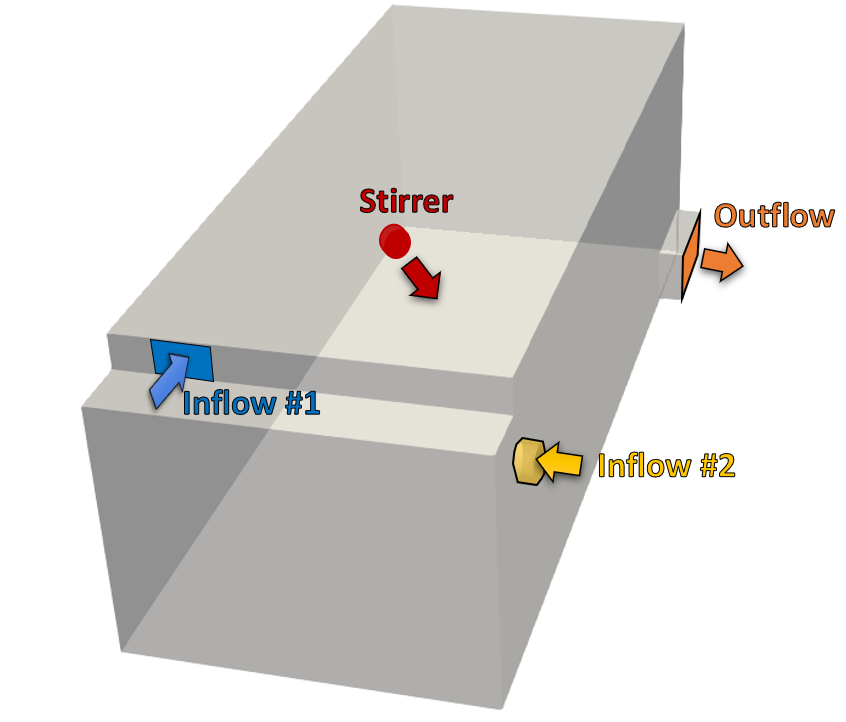"}
    \caption{Geometry of the reactor under study. Arrows represent the intended direction of the flow in the different areas of interest.}
    \label{fig:tank}
\end{figure}

Hence, our research is conducted leveraging the geometry of a full-scale homogenization tank used in a wide variety of industrial setups. Figure~\ref{fig:tank} depicts the geometry of the tank and the location of its areas of interest which are not simple walls and can be configured.
As it is shown, the tank is composed of a tank of $15 m$ length, $6 m$ width, and $5 m$ depth.
The flow enters the tank through two different locations \textit{inflow~\#1} and \textit{inflow~\#2} and it leaves the reactor through the \textit{outflow}.
Inside the tank, there is also a stirrer that is responsible for enhancing the mixing within the flow in order for the liquid or aqueous mixture to not settle at its bottom and continue its journey to the tank outlet. Depending on a set of different values, like the velocity of the impeller or the input flows of the tank, lower or higher velocities can be obtained inside the facility which can lead to affect the performance of the following operations. Therefore, analyzing a wide variety of cases of operation is useful to optimize the performance of the installation.

Notice that this study case is limited to evaluating a predictive model devised for and its performance when training it and making inferences with different resource configurations.
Thus, specific details about the tank are beyond the scope of this paper.
Nevertheless, similar tanks have been widely studied in other works where further applicable information can be found in~\cite{Climent2018, Iserte2021}.

This manuscript investigates the performance and different strategies of learning the AI model based on a data-driven approach. 
Then we provide interaction between AI and CFD solver for much faster analysis and reduced cost of trial \& error experiments. 
The scope of this research includes quasi-steady state simulations, which use an iterative scheme to progress to convergence. 
Quasi-steady state models perform a mass and energy balance of a process in an equilibrium state, independent of time \cite{Bhatt}. 
In other words, it is assumed that a solver calculates a set of timesteps to achieve the convergence state of the simulated phenomenon. Whence, the devised method is responsible for predicting the subsequent state with the AI model based on former timesteps generated by the CFD solver.

The most relevant contributions of this work are:
\begin{itemize}
\item an AI-based method that can be integrated with a CFD solver and allows users to predict the evolution of simulations based on initial timesteps generated by the CFD solver;
\item a recursive neural network (RNN) that predicts future states in a flow time series;
\item a performance analysis of different distributed training methods considering Horovod and Tensorflow \texttt{MultiworkerStrategy}; and
\item a performance comparison between CPU and GPU clusters for training and evaluating the devised predictive model.
\end{itemize}

The rest of the paper is structured as follows:
Section~\nameref{sec:materials} details the methods for simulating the flow, the RNN design, and validation, and introduces the distributed training techniques.
Section~\nameref{sec:perf} presents the performance evaluation of the training techniques.
Section~\nameref{sec:back} overviews the related work and research efforts where the presented study could be valuable.
The manuscript concludes in Section~\nameref{sec:conclusions} where future work is also included. 

\section{Materials and Methods}\label{sec:materials}
This section describes the CFD model that resolves the hydrodynamic inside the tank, as well as the predictive model, and how data is generated and processed.
Distributed training techniques are also presented in this section.

\subsection{CFD Simulation}
This point describes the CFD tools used in this work and a brief explanation of how they have been configured. All the CFD simulations have been performed with the software OpenFOAM\footnote{\url{http://www.openfoam.com}}.

In order to simulate the hydrodynamics in the tank, its geometry needs to be discretized into a mesh.
For this purpose, OpenFOAM tools \texttt{blockMesh}\footnote{\url{https://www.openfoam.com/documentation/user-guide/4-mesh-generation-and-conversion/4.3-mesh-generation-with-the-blockmesh-utility}} and \texttt{snappyHexMesh}\footnote{\url{https://www.openfoam.com/documentation/guides/latest/doc/guide-meshing-snappyhexmesh.html}} were leveraged.
The generated mesh corresponds to an octree with a maximum cell length of $0.213\,cm$, and refinement up to two levels for each boundary. 
Furthermore, the \textit{inflow~\#2} and \textit{outflow} surfaces are inflated with three layers.
As a result, the mesh domain is composed of $125,565$ cells.

The hydrodynamic numerical resolution inside the domain is performed with the OpenFOAM solver \texttt{pimpleFoam}\footnote{\url{https://www.openfoam.com/documentation/guides/latest/doc/guide-applications-solvers-incompressible-pimpleFoam.html}}, a large timestep transient solver for incompressible flow.
The flow equations are modeled using the Unsteady Reynolds-averaged Navier-Stokes (URANS) equations. 

In a nutshell, the fluid dynamic model is configured as follows:
\begin{itemize}
\item The kinematic viscosity is set to $0.000001\,m^2/s$, which is associated with the water properties.
\item For the boundary conditions, both, \textit{inflow~\#1} and \textit{inflow~\#2} surfaces are defined as a \texttt{flowRateInletVelocity} type condition with a constant volumetric flow rate. 
\item The \textit{outflow} surface is configured as \texttt{zeroGradient} type.
\item  The walls and top surfaces of the reactor count with the free-slip condition, while the remaining surfaces are specified as no-slip.
\item The timestep of the simulation is adjusted to the write interval, which corresponds to $10$ seconds of simulated time.
\end{itemize}

The flow is evolved until the second $4,201$ of the simulated time. In other words, there are stored 420 timesteps per executed simulation.

\subsection{Predictive Model}
This section describes the dataset generation, the predictive model design, its evaluation and validation, and the distributed training methods leveraged in the performance evaluation.

\subsubsection{Dataset}
Generating the dataset for the data-driven model requires performing a series of CFD simulations.
For this study, the execution has been limited to flow variations on the \textit{inflow~\#1} and \textit{inflow~\#2} boundaries. 
Concretely, the executed simulations correspond to 131 combinations of values from $0.1666$ to $0.3389$ for \textit{inflow \#1}; and values from $0.3333$ to $0.4443$ for \textit{inflow \#2}.

In this regard, the dataset is composed of 131 cases of 420 timesteps each.
In turn, every single timestep contains the information of 125,565 cells.
Although each cell in the domain can host several calculated metrics such as pressure or turbulence, the predictive model is focused on the three-dimensional metric of velocity.
The eventual dataset with shape $131 \times 420 \times 125,565 \times 3$ has a size of 38.6~GB in memory using the \texttt{float32} data type.

Before feeding the predictive model, each velocity dimension is normalized to have a distribution of mean zero and standard deviation of one.
Moreover, the dataset is split into train and test subsets.
For this purpose, cases are shuffled and 80\% of them (104 cases) are assigned to the training dataset, while the remaining (27 cases) are to the testing dataset.
Notice that 20\% of training cases (20 cases) are used for cross-validating the learning.

\subsubsection{Neural Network}
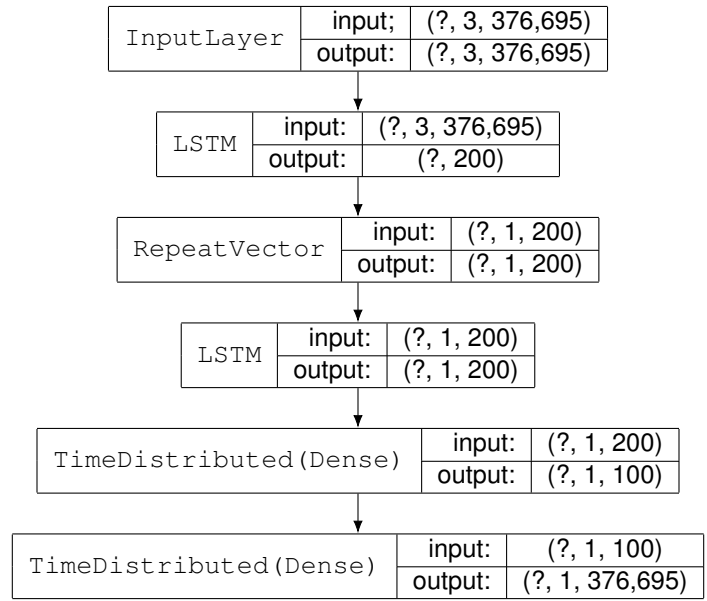
\begin{figure*}
\centering
\begin{tikzpicture}[font=\sffamily,>={Latex[bend]}]
 \matrix[matrix of nodes,nodes={anchor=center,inner sep=0pt,outer sep=0pt},
     ampersand replacement=\&,column sep=3em,row sep=1.5em](mat){
    \begin{tabular}{|l|r|c|}
     \hline
     \multirow{2}{*}{\texttt{InputLayer}}
     & input; & (?, 3, 376,695) \\
     \cline{2-3}
     & output: & (?, 3, 376,695)\\
     \hline
    \end{tabular} 
  \\
    \begin{tabular}{|l|r|c|}
     \hline
     \multirow{2}{*}{\texttt{LSTM}}
     & input: & (?, 3, 376,695) \\
     \cline{2-3}
     & output: & (?, 200)\\
     \hline
    \end{tabular} 
  \\
    \begin{tabular}{|l|r|c|}
     \hline
     \multirow{2}{*}{\texttt{RepeatVector}}
     & input: & (?, 1, 200)\\
     \cline{2-3}
     & output: & (?, 1, 200)\\
     \hline
    \end{tabular} 
  \\
    \begin{tabular}{|l|r|c|}
     \hline
     \multirow{2}{*}{\texttt{LSTM}}
     & input: & (?, 1, 200) \\
     \cline{2-3}
     & output: & (?, 1, 200)\\
     \hline
    \end{tabular} 
  \\
    \begin{tabular}{|l|r|c|}
     \hline
     \multirow{2}{*}{\texttt{TimeDistributed(Dense)}}
     & input: & (?, 1, 200) \\
     \cline{2-3}
     & output: & (?, 1, 100) \\
     \hline
    \end{tabular} 
  \\
    \begin{tabular}{|l|r|c|}
     \hline
     \multirow{2}{*}{\texttt{TimeDistributed(Dense)}}
     & input: & (?, 1, 100) \\
     \cline{2-3}
     & output: & (?, 1, 376,695) \\
     \hline
    \end{tabular} 
  \\
    };
\draw[->] (mat-1-1) -- (mat-2-1);
\draw[->] (mat-2-1) -- (mat-3-1);
\draw[->] (mat-3-1) -- (mat-4-1);
\draw[->] (mat-4-1) -- (mat-5-1);
\draw[->] (mat-5-1) -- (mat-6-1);
\end{tikzpicture}
\caption{RNN architecture. Next to the type of layer, the input and output data shapes of each layer are indicated.}\label{fig:dcnn}
\end{figure*}

The objective of the predictive model is to be capable of predicting time series given a short sequence of timesteps.
For this reason, an RNN able to retain temporal knowledge has been devised.
An RNN is a variation of feedforward neural networks whose neurons can be connected with cycles enabling neurons' outputs to be used in their own input creating an effect of memory~\cite{medsker1999recurrent}.
In this regard, the neural network presents temporal dynamic feature ideal for time series predictions.
Figure~\ref{fig:dcnn} depicts the layers architecture of the designed RNN.
Each layer in the figure describes the data shape of their input and output.
Notice that ``$?$'' symbol in the figure refers to the mini-batch size, which is not defined during the model design.

To begin with, the input layer expects three timesteps and the velocity values in each cell.
Since velocity is three-dimensional, the model expects to receive the velocity in a flat array of 376,695 values corresponding to $mesh\ cells \times velocity\ dimensions$.
As a result, the input shape is (3, 376,395).

Recurrence is leveraged with the Keras implementation of the Long Short-Term Memory (LSTM) layers~\cite{article}.
The proposed model is composed of two 200-unit LSTM layers.
The first one receives the input and transforms it accordingly to the given units to conform to the output space.
That output is repeated as many times as the desired predicted timesteps.
In this study, the predicted timesteps are limited to one.
At this point, each repeated vector follows its path composed of another LSTM, a 100-neurons fully-connected layer, and finally, another dense layer that outputs the initial velocity flat array.

The neurons within the hidden layers are activated with the rectified linear unit (ReLU) non-linear function which allows small positive gradients when the unit is not active~\cite{Maas2013}.

The model is compiled with the \textit{Adam} optimizer~\cite{Kingma2015} implemented in Keras to update weights and biases within the network. The optimizer is configured with a learning rate of $0.00025$.
The chosen loss function computes the mean of the absolute difference between labels and predictions (MAE):
\begin{equation}
MAE = \frac{1}{n}\sum_{i=1}^{n}{|y_i-y'_i|}.
\end{equation}

The output layer provides predictions with shapes (1, 376,395), corresponding respectively to the number of predicted timesteps and the flattened velocity within the domain cells.

Because of the hardware limitations (described in Section~\nameref{sec:perf}), the mini-batch size is set to 14.
The model has been trained until the validation loss no longer decreased during a period of ten epochs.
Through this process, the validation loss kept above the training loss showing good signs of overfitting prevention.
Eventually, the training was considered completed after 20 epochs, yielding a training loss in the magnitude of $10^{-2}$.

Furthermore, because of the dataset size, the model has to be fed with the help of a generator which not only arranges the mini-batches but also reshapes the data and the target accordingly to the network characteristics previously described.
In this regard, every requested mini-batch is provided by a generator based on the \textit{Sequence} Keras class.

Since the dataset is conformed by cases corresponding to the different configurations of the experiments, which in turn are composed of timesteps, the generator is responsible for rearranging the data to provide it seamlessly when it is requested by the model for training, validating, or evaluating.
Algorithm \ref{alg:generator} presents the generator methods.
The \texttt{Init} method initialize the data structures for the indexes and the samples.
With \texttt{Get\_item} a mini-batch of data through the dataset is yielded. The timesteps within the mini-batch are arranged as three observations and one target.
Finally, after each epoch of the training, indexes are shuffled (method \texttt{On\_epoch\_end}).

\begin{algorithm*}[t]
\caption{Methods of the \textit{Generator}}\label{alg:generator}
\begin{algorithmic}

\Function{Init}{}
\State $Indexes \gets \ array\ of\ samples\ per\ case\ (for\ these\ scenario\ all\ the\ cases\ have\ the\ same\ number\ of\ samples).$
\State $Samples \gets array\ of\ concatenated\ samples\ of\ all\ the\ cases.$
\State \Call{On\_epoch\_end}{}
\EndFunction

 \Function{Get\_item}{}
  \State $Calculate\ sample\ indexes\ of\ the\ minibatch.$
  \State $Arrange\ samples\ to\ the\ shape:\ (\#observation, \#target).$
  \State \Return minibatch
\EndFunction

 \Function{On\_epoch\_end}{}
\State $Shuffle\ indexes.$
\EndFunction
\end{algorithmic}
\end{algorithm*}

\subsubsection{Evaluation}
\begin{figure*}[tbp]
  \begin{center}
    \begin{subfigure}{.45\textwidth}
    \includegraphics[clip, width=0.99\linewidth]{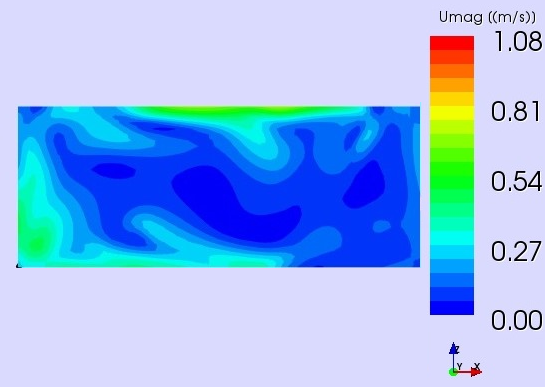}
    \caption{CFD}
    \end{subfigure}%
    \begin{subfigure}{.45\textwidth}
    \includegraphics[clip,width=0.99\linewidth]{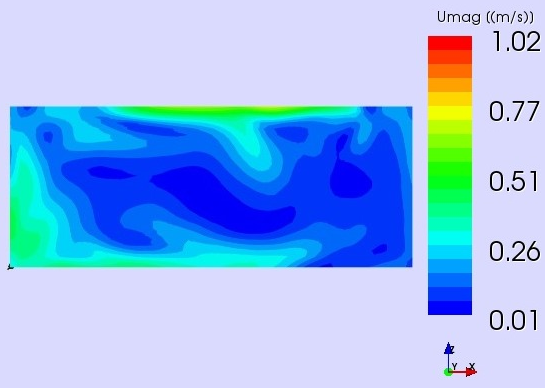}
    \caption{AI}
    \end{subfigure}%
    \caption{Contour plot of the velocity magnitude vector field ($U [m/s]$) using either the conventional CFD solver (a) or AI-accelerated approach (b) after ten timesteps.}
    \label{fig:U10}
  \end{center}
\end{figure*}

\begin{figure*}[tbp]
  \begin{center}
    \begin{subfigure}{.45\textwidth}
    \includegraphics[clip, width=0.99\linewidth]{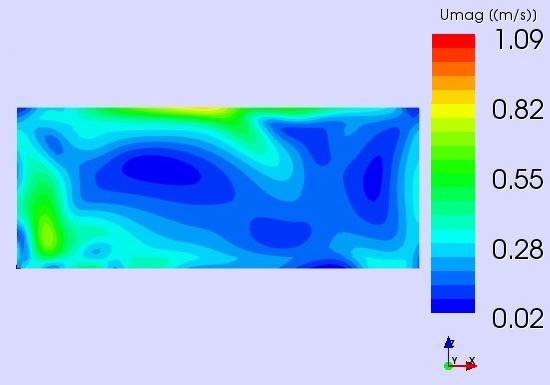}
    \caption{CFD}
    \end{subfigure}%
    \begin{subfigure}{.45\textwidth}
    \includegraphics[clip,width=0.99\linewidth]{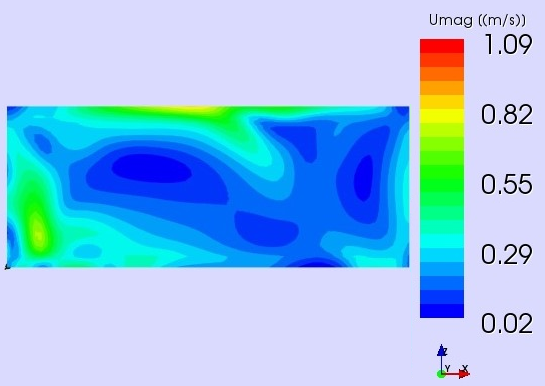}
    \caption{AI}
    \end{subfigure}%
    \caption{Contour plot of the velocity magnitude vector field ($U [m/s]$) using either the conventional CFD solver (a) or AI-accelerated approach (b) in the converged state.}
    \label{fig:U100}
  \end{center}
\end{figure*}

In this part of the study, the accuracy of the predictive model is evaluated. Since accuracy evaluation for the regression problems is not as obvious as for the classification process, a representative set of metrics have been chosen to validate the results. In the beginning, we compare the contour plots of the velocity magnitude created across the XZ cutting plane defined in the center point of the geometry. 
The results are shown in figures~\ref{fig:U10} and~\ref{fig:U100}, where the $10-th$, $20-th$, and the quasi-steady states calculated by the traditional CFD solver (left side) and AI-based method (right side), are showcased.

The results reveal high similarity, particularly for the values from the upper bound of the range.

\begin{figure*}[tbp]
  \begin{center}
    \begin{subfigure}{.45\textwidth}
    \includegraphics[clip,width=0.99\linewidth]{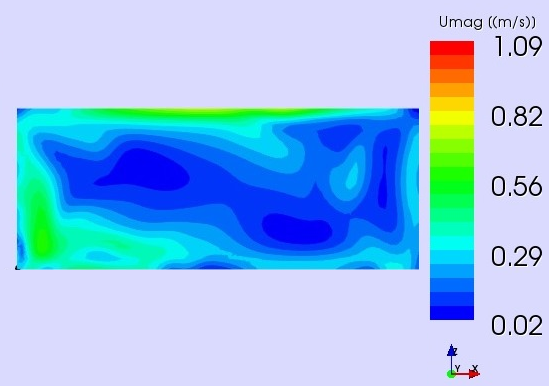}
    \caption{CFD}
    \end{subfigure}%
    \begin{subfigure}{.45\textwidth}
    \includegraphics[clip,width=0.99\linewidth]{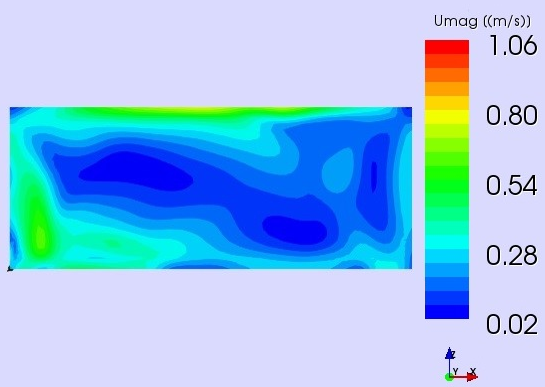}
    \caption{AI}
    \end{subfigure}%
    \caption{Contour plot of the velocity magnitude vector field ($U [m/s]$) using either the conventional CFD solver (a) or AI-accelerated approach (b) after 20 timesteps.}
    \label{fig:U20}
  \end{center}
\end{figure*}

Moreover, the accuracy is examined with a set of statistical metrics. The first two are correlation coefficients that measure the extent to which two variables tend to change together. These coefficients describe the strength and the direction of the relationship. Particularly, the coefficients selected are:
\begin{itemize}
\item Pearson correlation which estimates the linear relationship between two continuous variables; and
\item Spearman correlation which assesses the monotonic relationship between two continuous or ordinal variables. 
\end{itemize}

The Pearson correlation varies from $0.98$ for the $10-th$ timestep to $1.0$ for the converged state. 
The average Pearson correlation for all the timesteps is $0.99$. 
It shows a high degree of correlation between the CFD (computed by the solver) and AI (predicted) values. 
The Spearman correlation varies from $0.96$ to $1.0$ with the average value equal to $0.98$. It shows a strong monotonic association between the CFD and AI results.

The next statistical metric is the root mean square error (RMSE). 
It is the standard deviation of the residuals (prediction errors). 
Residuals are a measure of how far from the regression line data points are. 
RMSE varies from $0.041$ for the $10-th$ timestep to $0.005$ for the quasi-steady state. 
The average RMSE for all the validated timesteps is $0.023$. 
Based on these results, we conclude that the proposed AI models are well fit. 

The fourth method of accuracy examination is histogram equalization. 
In this method, histograms for the CFD solver and AI module results are created. They evaluate a numerical parameter that expresses how well two histograms match. 
The histogram comparison is made with the coefficient of determination, which is the percentage of the variance of the dependent variable predicted from the independent variable. 
The results vary from 87\% to 99\%, with an average accuracy of 98\%. 
All metrics are included in Table~\ref{tab:accFirst}.

\begin{table*}[!ht]
    \centering
    \begin{tabular}{r|c|c|c|c}
 Timestep \: & \: Pearson's coefficient. \: & \: Spearman's coefficient \: & \: RMSE \: & Histogram equalization \\\hline \hline
\: 10\ts{th}        & 0.984         & 0.969         & 0.041 & 97.3\% \\\hline
\: 20\ts{th}        & 0.995        & 0.993         & 0.023 & 98.0\%  \\\hline
\: Converged  & 1.000        & 1.000         & 0.005 & 99.5\%        \\
\end{tabular}
    \vspace{0.25cm}
\caption{Accuracy results with statistical metrics}
\label{tab:accFirst}
\end{table*}

\begin{figure}[h]
  \begin{center}
      \includegraphics[trim={0.55cm 0.5cm 0.25cm 0.25cm}, clip, width=\linewidth]{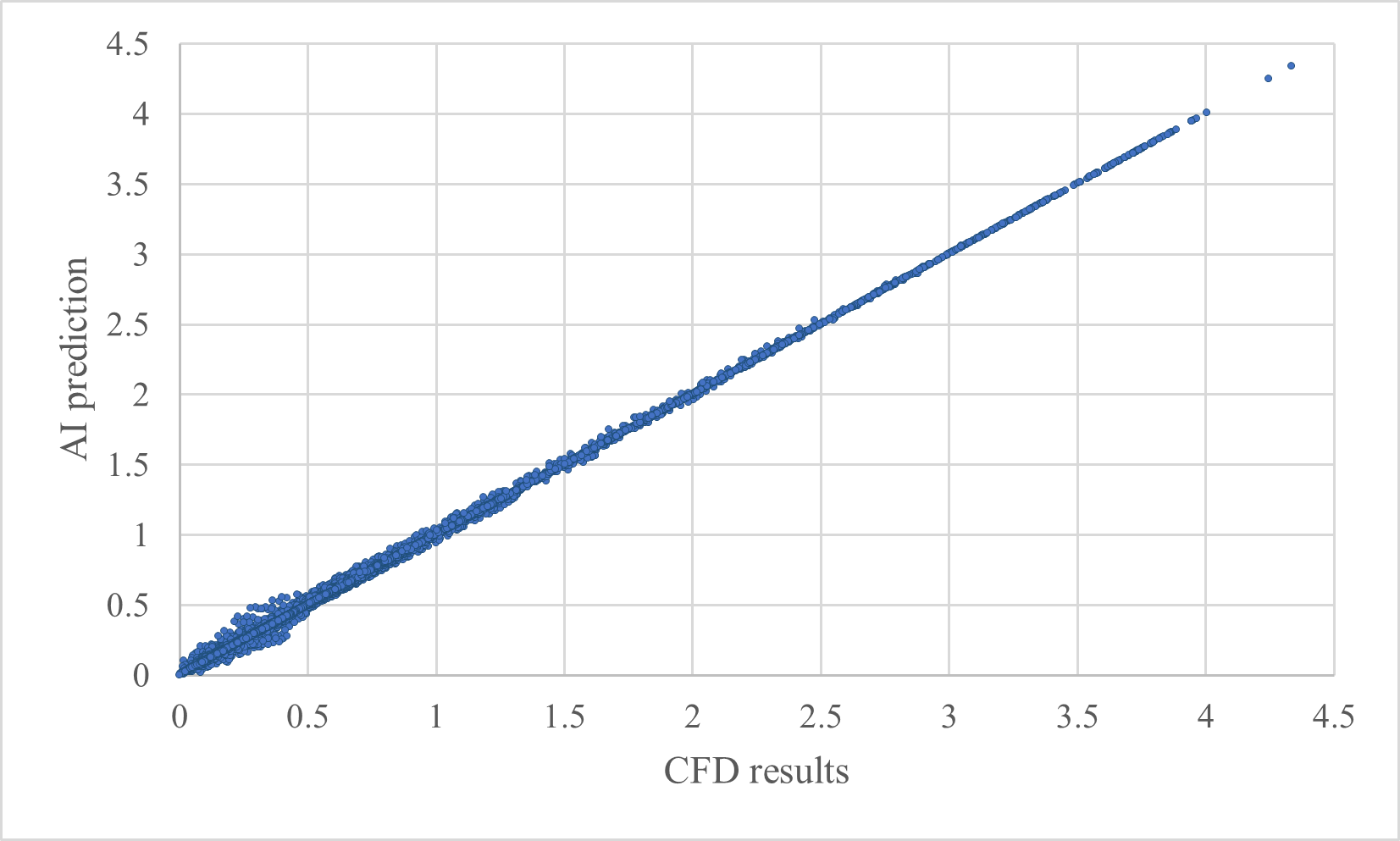}
    \caption{Comparison of simulation results for the conventional CFD solver and AI-accelerated approach.
    }
    \label{fig:priori}
  \end{center}
\end{figure}

Finally, a collective comparison of the results is performed, plotting a function $y(x)$, where $x$ depicts the results obtained from the CFD solver, while $y$ is the prediction. 
The results are shown in Figure~\ref{fig:priori}. 
The blue dots reveal the prediction uncertainty. 
In an ideal scenario, this test should be plotted as a straight line $y(x)=x$. 
The results confirm the previous conclusion that higher values (the most significant) are the most precisely predicted.

\subsection{Distributed Learning}\label{subsec:distributed}
Another contribution of this work has been to evaluate the performance of training the model using many accelerators distributed in several nodes.
For this purpose, this study leverages two different strategies which provide support for distributed training.
While the first strategy presented is implicitly supported by Tensorflow, the second is a framework for several deep learning libraries, among them, Tensorflow. 

\subsubsection{Tensorflow}
~\cite{10.5555/3026877.3026899} is an open-source library for artificial intelligence.
It is intended for the training and inference of deep neural networks.

Tensorflow provides native mechanisms of synchronous training across multiple replicas on one or more machines, particularly:
\begin{itemize}
    \item \textit{MirroredStategy\footnote{\url{https://www.tensorflow.org/api_docs/python/tf/distribute/MirroredStrategy}}:} 
    a variable created with this strategy is a \textit{MirroredVariable} replicated in all the involved GPUs, or CPUs if no GPUs are found. All the CPUs of a machine are treated as a single device (such as a GPU) and a thread per core is spawned for parallelism.
    
    \item \textit{Multiworker\footnote{\url{https://www.tensorflow.org/api_docs/python/tf/distribute/MultiWorkerMirroredStrategy}}:}
    this strategy complements \textit{MirroredStrategy} with the capability of communicating multiple nodes (\textit{workers}), each with its GPUs. For this purpose, it replicates all variables and computations to each local device. Multiple workers can work together using an all-reduce operation.
\end{itemize}

\subsubsection{Horovod}
~\cite{horovod} is a distributed learning framework implemented using the Message Passing Interface (MPI) model (although it supports other communication libraries such as Gloo) in an effort to reduce the configuration complexity introduced by \textit{MultiworkerStrategy} of Tensorflow to operate in computational clusters.
Horovod provides support for AI libraries such as PyTorch, MXNet, and TensorFlow.
Once a training script has been written for scaling, it can run on any number of GPUs and/or nodes without any further code changes. 

Horovod relies on the NVIDIA Collective Communication Library\footnote{\url{https://developer.nvidia.com/nccl}} (NCCL) to implement optimized multi-GPU and multi-node communications.
Horovod core is based on MPI primitives such as \textit{rank}, or \textit{all-reduce}.
In this regard, Horovod documentation informs that when using NCCL, performance will be similar to the original library.
However, thanks to the high degree of specialization of MPI in High-performance Computing (HPC) environments, especially in communications, users can expect that their only-CPU training runs faster.

\section{Performance Analysis}\label{sec:perf}
In this section, a thorough performance evaluation of the training, as well as the validation of the model, in different scenarios is performed.
Results presented in this section have been obtained with the following hardware:
\begin{itemize}
    \item CFD simulations are executed on Tirant III supercomputer. 
    Tirant III servers are composed of two sockets Intel Xeon SandyBridge E5-2670 @ 2.6GHz with a total of 16 threads, and 32GB of main memory.
    \item  The predictive model training and inference are performed on the CTE-Power cluster. 
    CTE-Power nodes are equipped each with two processors IBM Power9 8335-GTH @ 2.4GHz with a total of 160 threads, 512GB of main memory, and four GPU NVIDIA V100 with 16GB HBM2.
    \review{
    GPUs within a node are connected via NVLink 2.0 @ 25Gb/s. Concretely, GPUs 0-1, in each host, are connected traversing a bonded set of 3 NVLinks and PCIe bridges within the NUMA node of CPUs 0-79. Correspondingly, GPUs 2-3, are connected traversing their 3 NVLinks and the PCIe bridges within the NUMA node of CPUs 80-159. Each NUMA node is associated with a different Network Interface Card (NIC).
    }
    Nodes are interconnected via a single Port Mellanox EDR (25Gb/s).
\end{itemize}

Notice that with CTE-Power not only the neural network training but also the CFD simulation could have been performed. Nevertheless, we opted for using Tirant III for the CFD simulations since the solvers are not implemented for GPU architectures.
It would have been translated into a lot of wasted GPU time during their executions, and many users could have seen their research impaired.

Regarding the software stack, CFD simulations have been performed in Tirant III with OpenFOAM v2006 and Intel MPI 2018 Update 3.
While the predictive model has been developed on CTE-Power with Python~3.7.4, CUDA~10.2.89, Keras~2.4, Tensorflow-gpu~2.3, Horovod~0.20.3, and OpenMPI~4.0.1.

CFD simulations take an average of 9,000 seconds running on a 16-core node of Tirant III.
Moreover, for each of the 131 simulated cases, results have to be post-processed to extract and adapt data into a format readable by the predictive model. This postprocessing takes around 2,700 seconds in a single core.

The following showcased times correspond to exclusively the training time, not to the program initialization or the data loading.
The time of each configuration has been measured using the average of three executions.

In order to assess the reproducibility of the results, the random seed for initializing training variables has been fixed to a constant value.

\subsection{Only CPU Training}\label{subsec:cpu_train}
As a baseline, an initial study on training only with CTE-Power CPUs is performed on up to four nodes.
By default, Tensorflow and Horovod use all the cores with their threads, so the number of spawned processes is set to one per node.

To begin with, the model is fit with Tensorflow.
Using the traditional approach in a single node, the training time is 76,674 seconds.
When enabling distributed training with \textit{MultiworkerStrategy} a series of core dumps aroused because the Tensorflow-GPU version runs without GPUs, and that strategy does not support it.

On the contrary, execution times of the Horovod-based model fitting are compiled in Table~\ref{tab:trainCPU}.
The table also shows the speedup of Horovod compared to the Tensorflow baseline training time introduced before.

Notice that single-node execution performs better in Horovod since it is compiled explicitly for CPU, despite of Tensorflow-GPU installed in the CTE-Power machine that cannot leverage CPU optimizations.

\begin{table}[tbp]
    \small
    \centering
    \begin{tabular}{l|c|c|c|c}
        \# Nodes  & 1 & 2 & 3 & 4 \\\hline
        Time (s.) & 75,346 s. & 66,573 s.& 58,678 s.& 52,908 s.\\\hline
        Speedup   & 1.02x & 1.15x & 1.31x & 1.45x \\
    \end{tabular}
     \vspace{0.25cm}
    \caption{Horovod CPU training time and speedup compared with the single-node Tensorflow training.}
    \label{tab:trainCPU}
\end{table}

Although the speedup increases with the number of nodes, it is far from being linear and too high compared with the GPU-enabled configuration showcased in the next section.

\subsection{GPU-enabled Training}
In this subsection, the RNN distributed training on GPUs is studied.
As it was described in Section~\nameref{subsec:distributed}, the distributed training of the model in TensorFlow could be performed either with its native tools (\textit{MultiworkerStrategy}) or with the framework Horovod.
For this reason, the training was implemented in both alternatives.

Firstly, distributed training in Tensorflow has been evaluated.
Table~\ref{tab:trainGPU-mw} contains the training results of \textit{MultiworkerStrategy} for a different number of nodes involved.
In this evaluation, only a GPU per node has been leveraged.

\begin{table}[tbp]
\small
    \centering
    \begin{tabular}{l|c|c|c|c}
        \# Nodes  & 1 & 2 & 3 & 4 \\\hline
        Time (s.) & 10,803 s.& 43,288 s.& 38,420 s.& 32,928 s.\\\hline
        Speedup   & - & 0.25x & 0.28x & 0.33x \\
    \end{tabular}
     \vspace{0.25cm}
    \caption{\textit{MultiworkerStrategy} GPU training time and speedup}
    \label{tab:trainGPU-mw}
\end{table}

However, distributing the workload in several nodes is translated into performance degradation.
This is a known issue in CTE-Power for TCP/socket-based networking communications which are reported to have half the bandwidth.
Apart from these issues, the model training still scales along with the number of nodes.
The study with \textit{MultiworkerStrategy} is not continued with more GPUs per node due to the high times obtained in this stage.
In this regard, henceforth, the results of distributed configurations correspond to Horovod with MPI, compiled to use Infiniband over the Mellanox EDR ports installed in the nodes.

Table~\ref{tab:trainGPUspeedup} contains the speedups of the different configurations of nodes and GPUs compared with Tensorflow on a single GPU (second row and column).
The results of \textit{MirroredStrategy} for several GPUs in a single node are placed in the second row of the table.
The rest of the rows contains the results using Horovod with a different number of nodes and GPUs.

\begin{table}[tbp]
\small
    \centering
    \begin{tabular}{l|r|r|r|r}
        ~ & 1 GPU & 2 GPUs & 3 GPUs & 4 GPUs  \\ \hline \hline
        TF \texttt{mirrored} & 1.00x & 1.61x & 2.05x & 2.41x  \\ \hline
        Horovod 1 node &   0.99x & 1.60x & 2.08x & 2.40x  \\ \hline
        Horovod 2 nodes &  1.58x & 2.25x & 2.70x & 3.01x  \\ \hline
        Horovod 3 nodes &  1.98x & 2.64x & 2.96x & 3.22x  \\ \hline
        Horovod 4 nodes &  2.20x & 2.84x & 3.40x & 3.31x  \\
    \end{tabular}
     \vspace{0.25cm}
    \caption{GPU training speedup}
    \label{tab:trainGPUspeedup}
\end{table}

In the previous tables, we can also appreciate that for the same number of processes depending on how they are distributed among the underlying hardware, times vary.
The tables reveal that the training operation scales with the amount of leveraged resources.
From those results, we can calculate the negative effect of the communication network in different processes layout.
Notice that the amount of spawned processes corresponds to the expression $\#Processes = \#Nodes \times \#GPUs$.

Table~\ref{tab:network} compiles the communication overhead for the different layouts for the same number of processes.
Each cell contains the variation in the percentage of time compared with the symmetric cell on the other side of the diagonal. For instance, two processes in a node using two GPUs run 0.89\% faster than the configuration of two nodes using a GPU in each one, which needs 0.9\% more time to complete.

\begin{table}[tbp]
\small
    \centering
    \begin{tabular}{l|r|r|r|r}
        ~ & 1 GPU & 2 GPUs & 3 GPUs & 4 GPUs  \\ \hline \hline
        1 node  &       - &  0.89\% &  4.55\% &   8.64\%  \\ \hline
        2 nodes & -0.90\% &       - & 2.22\% &   5.73\%  \\ \hline
        3 nodes & -4.76\% &  -2.27\% &       - & -5.76\%  \\ \hline
        4 nodes & -9.46\% & -6.08\% & 5.45\% & -  \\
    \end{tabular}
     \vspace{0.25cm}
    \caption{Communication network effect in Horovod on execution times comparing the same number of processes distributed in different layouts.}
    \label{tab:network}
\end{table}

Theoretically, configurations on the upper triangle should run faster (positive percentages) than in the lower triangle (negative percentages), since fewer nodes involve less communication.
However, the 12-GPU case reveals that running on four nodes (three GPUs each) is faster than on three nodes (four GPUs each).

\begin{figure*}[h]
  \begin{center}
    \includegraphics[trim={0.45cm 0.4cm 1.5cm 0.2cm}, clip, width=0.99\textwidth]{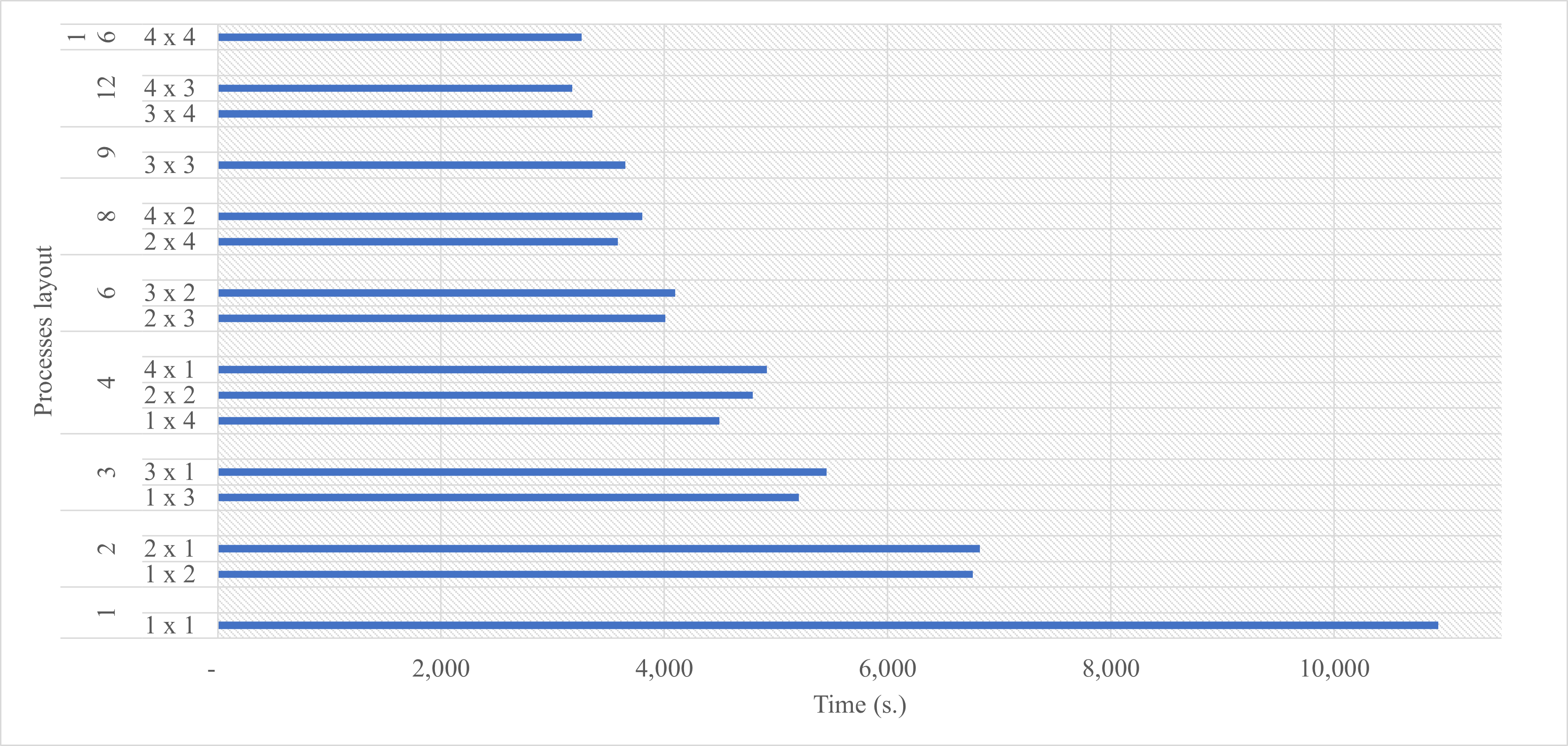}
    \caption{Horovod deep neural network training time for different processes configurations and layouts (nodes x processes).}
    \label{fig:layout}
  \end{center}
\end{figure*}

In more detail, Figure~\ref{fig:layout} represents the execution time of layouts for a different number of processes. In the y-axis, the number multiplying on the left is the number of nodes and the number of GPUs on the right.

The theoretical trend of reducing execution time with a lower number of nodes for the same amount of processes is followed, except for the previously named case of twelve processes.

\begin{figure*}[tbp]
  \begin{center}
    \includegraphics[trim={0.25cm 0.1cm 0.25cm 0.2cm}, clip, width=0.99\textwidth]{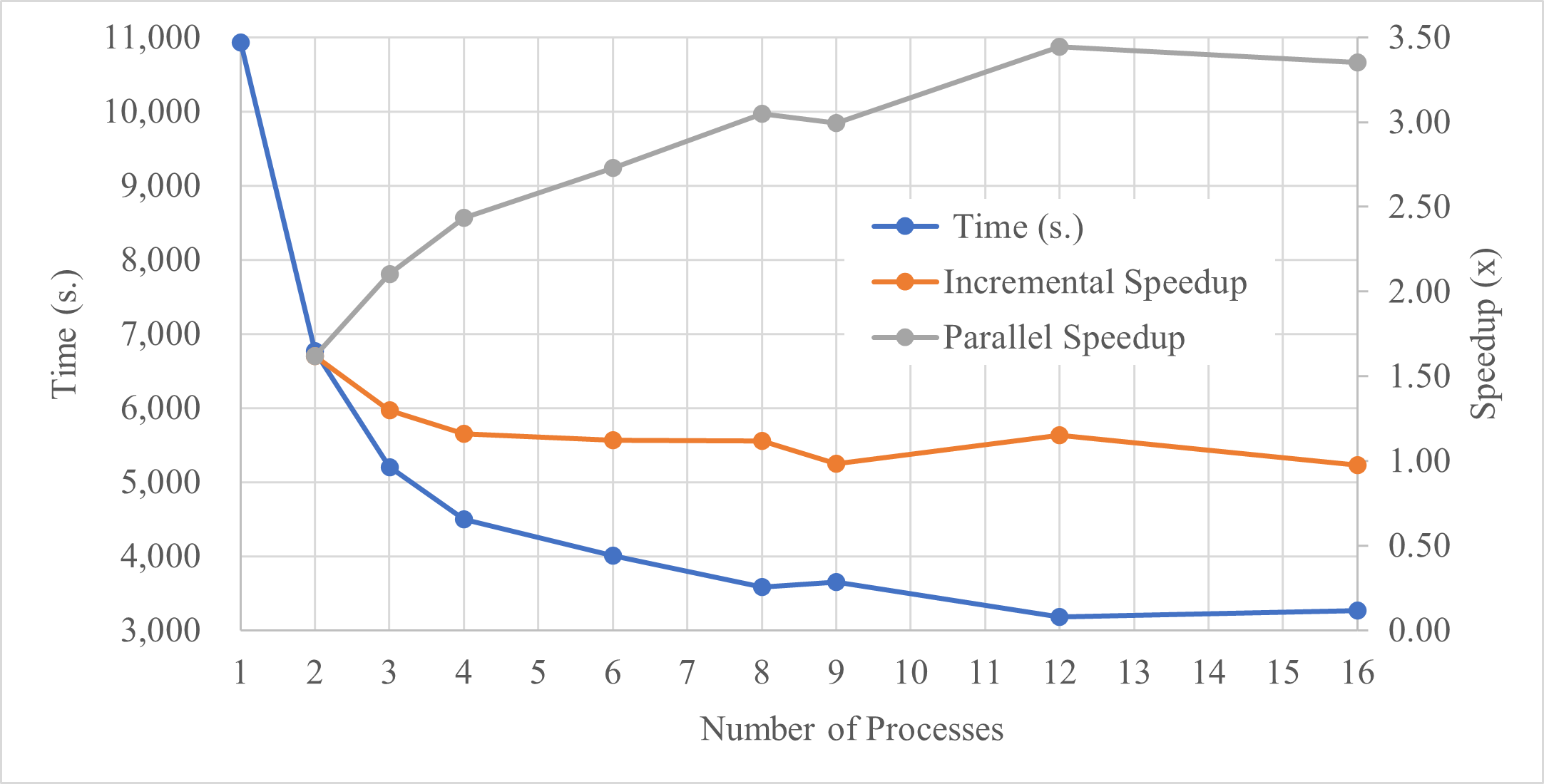}
    \caption{Training time and speedups for different process configurations in Horovod.}
    \label{fig:scal}
  \end{center}
\end{figure*}

In order to study the performance scalability, Figure~\ref{fig:scal} represents the training time for the lowest time layout of each process configuration.
Furthermore, times are associated with the incremental speedup (the time of a configuration divided by the time of the immediately previous one), and the parallel speedup (the time of a configuration divided by the 1-process configuration).

In the chart, can easily be spotted the best performance configuration with the lowest time and the highest parallel speedup. It corresponds to the 12-process configuration.
In this regard, it looks like because of the nature of the problem it fits best in this configuration when using four nodes.

Moreover, depending on our needs, we can determine a ``sweet spot'' configuration that achieves a balance between execution time and resources utilized.
The ``sweet spot'' for this study may be set to the two-process configuration, where the speedup increment is above 1.5x. 
And it would probably be in shared environments, however, since all the experiments have been performed using exclusive nodes, the best balance between time and resources can be defined as four processes running in the same node.
This configuration shows an incremental speedup of 1.16x and a parallel speedup of 2.43x.

\subsection{Inference}
The inference time of the evaluation dataset has also been measured.
Particularly, Table~\ref{tab:inference} showcases the prediction times for all the samples in the test dataset for one node, and in the case of the GPU-enabled inference, one GPU.

\begin{table}[tbp]
    \centering
    \begin{tabular}{c|c|c}
        CPU time & GPU time & GPU Speedup  \\ \hline
        3,136 s.& 1,014 s.& 3.09x  \\
    \end{tabular}
     \vspace{0.25cm}
    \caption{Comparison of inference results.}
    \label{tab:inference}
\end{table}

\section{Related Work}\label{sec:back}

The adaptation of CFD codes to hardware architectures exploring modern compute accelerators is an important issue in the environment of constantly growing needs for computing power. The acceleration of HPC simulations with AI becomes a popular topic these days since it offers much more efficiency than traditional hand-tuning.
In other words, AI methods in CFD simulations can reduce the time-to-solution while providing acceptable accuracy~\cite{vinuesa2021potential}.

\review{
Most of the technological advances in data-driven CFD simulations do not pay attention to scalability:
\begin{itemize}
\item In \cite{ROJ_2021}, the authors proposed a method for accelerating a chemical-mixing simulation with an AI model. This work shows that AI generates results for a MixIT tool with 92\% accuracy and speedup to 10x compared with the MixIT tool. The performance results are achieved on a single CPU/GPU node.
\item The authors in~\cite{Xiao2019} devised a Non-Intrusive Reduced Order Model (NIROM) combining Proper Orthogonal Decomposition (POD) and machine learning techniques. 
Their results show that the model is capable of predicting the evolution of a turbulent flow in a quasi-steady state. 
However, this solution is based on a limited dataset with a reduced number of scenarios trained in a 10-core machine.
\item In \cite{KIM}, the authors present a novel generative model to synthesize CFD simulations from a set of reduced parameters. The authors show that linear functions are less efficient than their non-linear counterparts.
They evaluate it with a single GTX 1080 Nvidia GPU.
\end{itemize}
}
\review{
Other works present surrogate CFD-DL models in which some stages of timesteps are predicted and others calculated.
In this regard, if we compare traditional CFD simulations with their AI-assisted counterparts, results in similar errors in a fraction of the execution time:
\begin{itemize}
    \item A deep learning framework coupled with a physical simulator for RANS models, named CFDNet, is presented in~\cite{Obiols-Sales2020}. CFDNet is intended for more generalist simulations, and it always applies POD methods~\cite{pod} to reduce the data dimensionality before training the predictive model. In their work, the authors compare several CFD simulations with their coupled model results obtaining similar errors in a reduced time.
\item A hybrid CFD-AI solver is proposed in~\cite{Iserte}. The solver alternates stages of CFD simulation with predictions made by a neural network previously trained with a given model. In this case, the neural network learns spatial-temporal features of the data leveraging convolutional kernels. The novel technique can significantly reduce the simulation time of a transient flow. The basic difference in our approach is to use an RNN-based model rather than a CNN-based one. 
\item J. Thompson et al.~\cite{THOMPSON} propose a data-driven approach that utilizes deep learning to obtain fast and highly realistic simulations. They use a CNN with a highly tailored architecture to solve the linear system. The key contribution is to incorporate loss training information from multiple timesteps and perform various forms of data augmentation.
\end{itemize}
}
\review{
Our research is devoted to investigating the training performance rather than inferencing and interaction with CFD solver.
}
\review{
Finally, related to scalability evaluation, we find:
\begin{itemize}
\item In~\cite{Ramirez} the authors leverage Tensorflow, the Math Kernel Library (MKL), and Horovod to accelerate the training of neural networks based on convolutional architecture such as AlexNet or ResNet-50.
The authors evaluate only-CPU distributed training in different clusters such as Marenostrum IV, CTE-Power9, and the ARM-based Dibona.
The comparison among architectures presents a scalability variation between 70\% and 99\% when using up to 16~nodes.
\item Awan et al.~\cite{Awan} evaluate the training of AlexNet in a Horovod implementation over several GPUs distributed across a cluster.
They showcase an almost perfect speedup of up to 256~GPUs.
\item In~\cite{Malik} the authors discuss how different communications protocols such as MPI or NCCL2 can differ in performance. A state--of--the--art protocol with near-ideal scaling is also presented.
\item Tensorflow \texttt{MirroredStrategy} with four GPUs is evaluated in~\cite{Quang}.
For a neural network trained with Mnist or Cifar-10 datasets, the maximum achieved performance is 2x.
\end{itemize}
}
\review{
This is the first work that widely evaluates the training scalability of a CFD time-series predictor.
This work has evaluated a data-driven CFD model that unlike classification models 
are based on a few layers and large amounts of data which increases the communication traffic.
}

\section{Conclusion and Future Work}\label{sec:conclusions}
This work investigates the distributed training of the RNN model for data-driven CFD simulation. 
Two methods including Horovod distribution and Tensorflow \textit{MirroredStrategy} and \textit{MultiworkerStrategy}, are applied and studied. 

The proposed DL model is validated with an OpenFOAM solver and the results allow us to achieve an accuracy of around 99\% based on the histogram equalization of the converged state when predicting single timesteps.

The performance of the training and inference stages between the CPU and GPU configurations considering different layouts of processes have been also analyzed. 
The research shows that when comparing performances between no-GPU and GPU scenarios, the training time with a GPU reduces the training time by 7.1x.
If the CPU time is compared to the maximum performance configuration time of four nodes and three GPUs per node, the speedup is 24.49x.
On the contrary, in the inference stage which is expected to be less concurrent, with a GPU the time decreases by 3.09x.

\review{
It is interesting to note that some of the low speedups obtained when using multiple GPUs are because having a small model and a large amount of data results in significant communication overhead and overhead of coordinating the computations across the GPUs, which outweighs the performance benefits of using multiple GPUs. This can lead to a low speedup and reduced scalability relative to the number of GPU devices.
A detailed analysis of the TensorFlow library and its implementation was outside the scope of this research. Although this research has provided a high-level overview of some of the potential reasons for low speedup when using multiple GPUs, further research and analysis may be needed to fully understand the behavior of TensorFlow in a multi-GPU environment and to identify the best practices for maximizing performance.
}

Besides the performance analysis, another interesting insight about the comparison of native distributed training methods of Tensorflow with Horovod is the user experience.
Horovod MPI-enabled leverages the network configuration of the cluster deployed for MPI.
In this regard, users can exploit all the network potential without further tuning.
Nevertheless, irrespective of the network communication issues, while with Horovod the environment is simply set up with a call to \texttt{init}, \textit{MultiworkerStrategy} needs to carefully define the cluster configuration of the nodes and GPUs involved in the execution.

The model presented in this paper and the subsequent performance analysis can be extremely helpful in hybrid AI-CFD strategies of simulation based on surrogate models or co-simulation such as the presented in~\cite{Srivastava} or~\cite{Paul2019} where CFD models are complemented with deep learning surrogate models of the high computational demanding subdomains of the case.
\review{
Furthermore, the interest in time series predictions and data-driven CFD simulations, led to port the predictive model developed in this work to the Intelligence Processor Units (IPU) architecture to evaluate the training performance on other platforms~\cite{roy}. 
}

All the codes developed for this study are available in 
\url{https://github.com/AlejandroGB13/CFD\_AI}. The datasets can be provided on demand.


\begin{funding}
\ifdefined\ANON
\author{Funding omitted for anonymized reviews}
\else
Researcher Sergio Iserte was supported by the postdoctoral fellowship APOSTD/2020/026 from the Valencian Region Government and European Social Funds.
\fi
\end{funding}


\section*{Acknowledgements}
CTE-Power cluster of the Barcelona Supercomputing Center (BSC-CNS), and Tirant III cluster of the \textit{Servei d'Informàtica} of the University of Valencia (UV) were leveraged in this research.
\review{Authors want to thank the anonymous reviewers whose suggestions significantly improved the quality of this manuscript.}

\bibliographystyle{SageH}
\bibliography{bib}

\balance

\vspace{0.5cm}

\begin{wrapfigure}{l}{25mm}
  \includegraphics[width=1in,height=1in,clip,keepaspectratio]{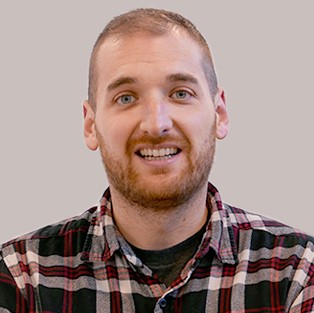}
\end{wrapfigure}\par
\textbf{Sergio Iserte}
Sergio Iserte holds the degrees of BS in Computer Engineering (2011), MS in Intelligent Systems (2014), and Ph.D. in Computer Science (2018) from Universitat Jaume I (UJI), Spain. Sergio Iserte is a senior researcher at Barcelona Supercomputing Center (BSC) in the Computer Science Department, and course instructor of the HPC subject at Universitat Oberta de Catalunya (UOC). He is currently involved in HPC projects related to parallel distributed computing, resource management, workload modeling, and deep learning for industrial applications.\par

\vspace{0.5cm}

 \begin{wrapfigure}{l}{25mm}
   \includegraphics[width=1in,height=1in,clip,keepaspectratio]{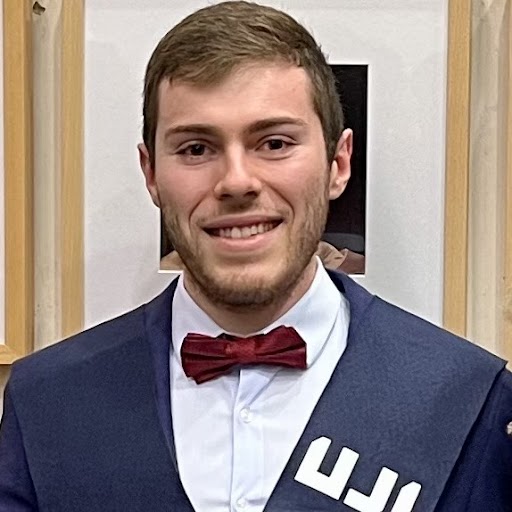}
 \end{wrapfigure}\par
 \textbf{Alejandro González-Barberá}
Alejandro González is a computer science graduate (2022) from Universitat Jaume I (UJI). Currently pursuing a Master's degree in Data Science at Universitat Oberta de Catalunya (UOC). Alejandro González is a junior researcher at Grupo Fluidos Multifásicos (GFM) at UJI. At the present time, his work focuses on the application of deep learning to industrial processes and water flow management.
\par

\vspace{0.5cm}

\begin{wrapfigure}{l}{25mm}
  \includegraphics[width=1in,height=1in,clip,keepaspectratio]{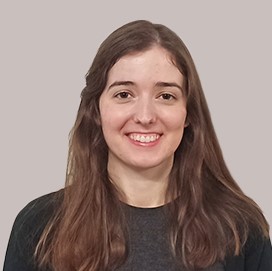}
\end{wrapfigure}\par
\textbf{Paloma Barreda}
Paloma Barreda holds the degree (2018) and MS in Industrial Engineering (2020) from Universitat Jaume I (UJI), Spain. Currently, she is an innovation agent of UJI through the Line: Promotion of talent of the Valencian Innovation Agency (AVI). In addition, she is also a researcher in the field of CFD modelling applied to WWTF (wastewater treatment facilities) in the Multiphase Fluids Group of the same institution where she is working on her PhD thesis.\par

 \vspace{0.5cm}

 \begin{wrapfigure}{l}{25mm}
  \includegraphics[width=1in,height=1in,clip,keepaspectratio]{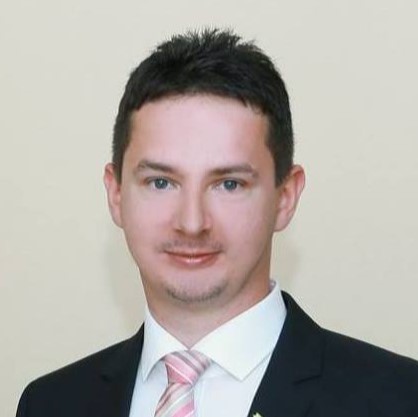}
\end{wrapfigure}\par
\textbf{Krzysztof Rojek}
Krzysztof Rojek is a researcher in the field of artificial intelligence (AI) and high-performance computing (HPC). He holds a Ph.D. and D.Sc.-Habilitation degree from the Czestochowa University of Technology. Throughout his career, he has made contributions to the advancement of these disciplines, publishing more than 30 papers and articles on topics ranging from machine learning algorithms to parallel computing. He is a professor at the Czestochowa University of Technology, where he has been teaching and conducting research in the fields of AI and HPC. In addition to his academic work, he is also the Chief Technology Officer (CTO) at byteLAKE, a company specializing in cutting-edge technology solutions.\par

\end{document}